\documentclass{appolbb80}
\usepackage{graphicx}
\usepackage{mathtools} 		
\usepackage{amssymb,amsmath,bbold}
\usepackage{float}	 % fixed positioning of figures (spezifier H)
\usepackage{color}
\usepackage{hyperref}
\newcommand{\nc}{\newcommand} %%%%%% Just to make it easier to write new commands
\nc{\req}[1]{Eq.\,\ref{#1}} 
\nc{\rf}[1]{Fig.~\ref{#1}} 
\nc{\Th}{\ensuremath{T_\mathrm{H}\,}}
\begin{document}
% \eqsec % uncomment this line to get equations numbered by (sec.num)
\title{\uppercase{Charting the future frontier(s) of particle production}%
%\thanks{Dedicated to Andrzej Bia\l{}as in honour of his 80th birthday}%
}
\author{Jan Rafelski
\address{Department of Physics, The University of Arizona\\
1118 E. Fourth Street, P.O. Box 210081, Tucson, AZ 885721, USA}
}
\maketitle
\begin{abstract}\vskip -1.1cm{\centering \hspace*{-1cm}(Received June 14, 2016)\\[0.3cm]
 \hspace*{1cm} {\em Dedicated to Andrzej Bia\l{}as in honour of his 80th birthday}}\vskip 0.2cm
This short note describes the long collaborative effort between Arizona and Krak\'ow, showing some of the key strangeness signatures of quark-gluon plasma. It further presents an annotated catalog of foundational questions defining the research frontiers which I believe can be addressed in the foreseeable future in the context of relativistic heavy ion collision experiments. The list includes topics that are specific to the field, and ventures towards the known-to-be-unknown that may have a better chance with ions as compared to elementary interactions.\\[0.2cm] To appear in Acta Phys. Pol. B \url{doi:10.5506/APhysPolB.47.1977}
\end{abstract}
%\PACS{12.38.Mh,12.40.Ee,25.75.-q}
% 12.38.Mh Quark-gluon plasma (see also 25.75.Nq Quark deconfinement, quark-gluon plasma production and phase transitions in relativistic heavy ion collisions; see also 21.65.Qr Quark matter)
%12.40.Ee Statistical models
%25.75.-q Relativistic heavy-ion collisions (collisions induced by light ions studied to calibrate relativistic heavy-ion collisions should be classified under both 25.75.-q and sections 13 or 25 appropriate to the light ions)
%%%%%%%%%%
\thispagestyle{empty}
%%%%%%%%%%%%%%%%%%%%%%
\section{The beginning}\label{Sec1}
Some 70 years ago the development of relativistic particle accelerators heralded a new era of laboratory-based systematic exploration and study of elementary particle interactions. For the past 50+ years Andrzej Bia\l{}as participated in this endeavor, entering the field at the time when two pillars of our present day understanding, the quark content of hadrons, and the Hagedorn statistical model of hadron production, were discovered.

The outcomes of this long quest are on one hand the standard model (SM) of particle physics, and on another, the discovery of the primordial deconfined quark-gluon plasma (QGP). These two foundational insights arose in the context of our understanding of the models of particle production and more specifically, the in-depth understanding of strong interaction processes. To this point we recall that in the context of SM discovery we track decay products of {\it e.g.\/} the Higgs particle in the dense cloud of newly formed strongly interacting particles. In the context of QGP we need to understand the gas cloud of hadrons into which QGP decays and hadronizes. Hadrons are always all we see at the end. They are the messengers and we must learn to decipher the message.

I wish to add here a few personal reminiscences. At first our scientific world lines intersected without us meeting personally: during preparation for my move to Arizona I was invited to lecture at Zakopane Summer school in 1986, and was featured in the poster that ornamented my wall for the following decades. Alas, I was not able to resolve out of Cape Town infrastructure challenges and the meeting happened without me. However, I tracked the research program of the organizer, and became an early admirer of intermitency in multiparticle poduction~\cite{Bialas:1988wc}. When I invited Andrzej in 1988 to report on this new topic in Tucson at the \lq\lq Hadronic Matter in Collision 1988\rq\rq\ meeting, as the tide of times turned, it was he who could not come and instead Rob Peschanski accepted the joint invitation. 

However, we could not escape a meeting for much longer. In October 1988 after 24 years I returned to Poland and saw my former home torn by civil war of disobedience. In all the disorder there was an almost working institute (all but the restrooms, as I was told, these were filled with special commune-fumes). I enjoyed many interesting conversations as well as a memorable evening at Wierzynek Restaurant. The fact is that we did have many things in common in science and personal backgrounds though it must go without saying, there are rarely two people of such opposite character! 

Years and decades passed; we saw each other semi-regularly at the Zakopane Summer school venue. However, some special events from the period come to mind:\\
{\bf a)} Around 1995 I was preparing a long review article on strangeness as a signature of QGP -- this was a time where a bound printed copy placed into hands of the \lq consumer\rq\ was important, and when the cost of \lq reprints\rq\ was where the cost of publication was hidden. This work found its path into the pages of Acta Physical Polonica B~\cite{Rafelski:1996hf}, filling its own issue and many, many reprints were distributed.\\
{\bf b)} An unintended outcome of this exercise was that Andrzej, who may have been the referee, became interested in the interpretation of strange antibaryon experimental results. Considering ratios of strange antibaryons measured at CERN-SPS, Andrzej in 1998 saw in the early data evidence for a new state of matter~\cite{Bialas:1998ea}, agreeing with our insights~\cite{Rafelski:1996hf}.\\
{\bf c)} To fill the need to create a public program allowing the experimental groups the study of the hadronization process quantifying their own results in a full description, the Krak\'ow IFJ and Tucson groups joined forces in 2002. Andrzej was instrumental in helping to build trust and collaborative spirit, and the outcome is the SHARE suite of programs, fully vetted by both groups~\cite{Torrieri:2004zz}.\\
{\bf d)} Visiting Krak\'ow a dozen years ago I came to Andrzej\rq s office for a chat which turned more serious. When the outcome was written up, Andrzej would not let me loose; in 2005 we finally became coauthors~\cite{Bialas:2005ij}.\\
{\bf e)} The picture of Andrzej, \rf{AB_SQM} was taken in 2011 on occasion of the Strangeness in Quark Matter (SQM) 2011 meeting where we had a good time celebrating my birthday, see \rf{SQM2011ABJR}.
%%%%%%%%%%%%%%%%%%%%%%%%%%%%%
\begin{figure}%[bht]
\centerline{%
\includegraphics[width=12.5cm]{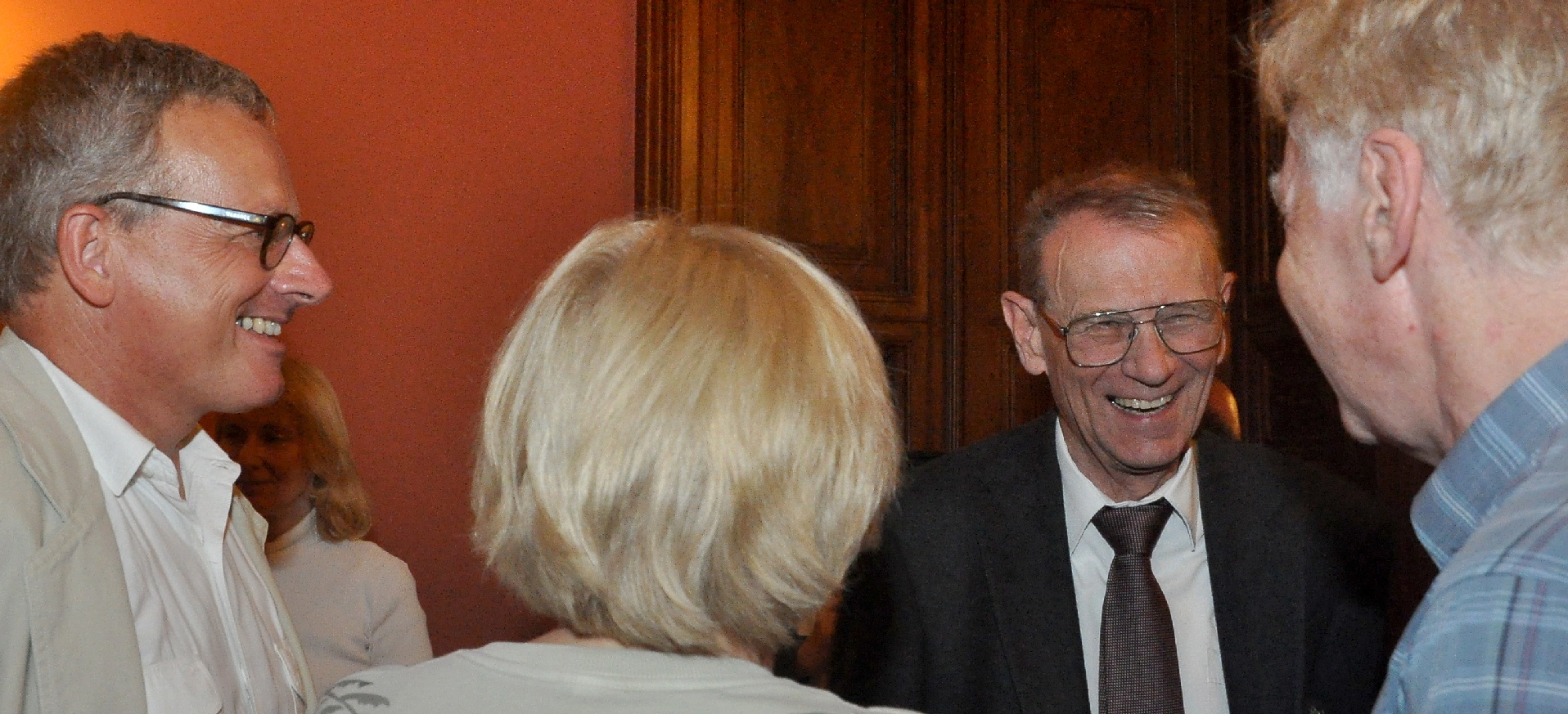}}
\caption{Andrzej Bia\l{}as (off-center right) in conversation with Micha\l{}\ Prasza\l{}owicz (on left) and Ludwik and Boguta Turko. Hiding behind Micha\l{}\ is Victoria Grossack. Photo taken at Krak\'ow SQM2011 on September 18, 2011 by Maciej Rybczy\'nski.}
\label{AB_SQM}
\end{figure}
%%%%%%%%%%

%%%%%%%%%%%%%%%%%%%%%%%%%%%%%
\begin{figure}%[htb]
\centerline{%width=12.5cm
\includegraphics[width=\linewidth]{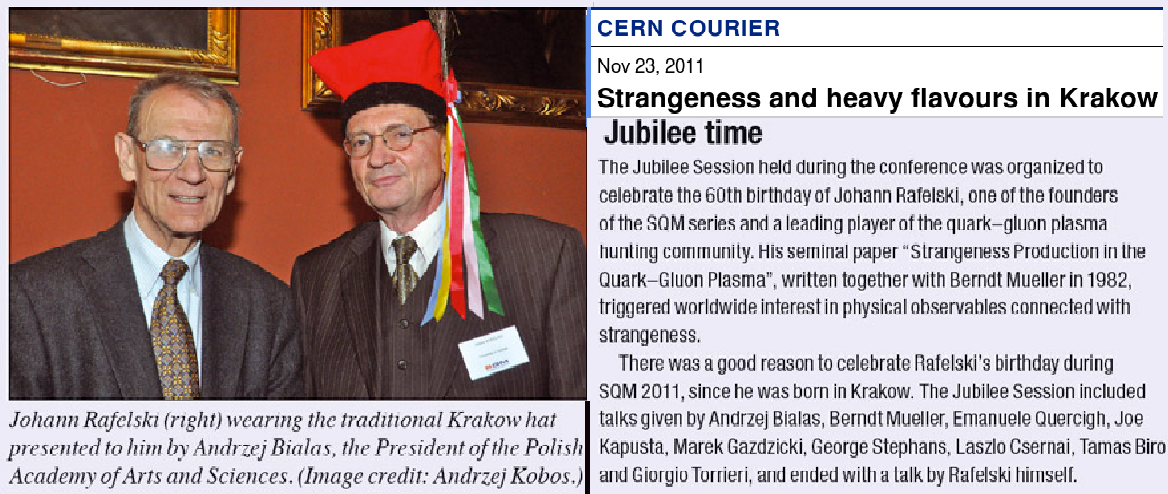}
}
\caption{SQM2011 meeting notice in the CERN Courier of December 2011 (partial).}
\label{SQM2011ABJR}
\end{figure}
%%%%%%%%%%%%%%%%%%%%%%%%%%%%%

Perhaps most important of all, for many decades we have shared our fascination about strong interactions, where we both see the challenge of the deeply unknown: quark confinement is a feature encoded into the properties of the empty space, the structured quantum vacuum. But really, are we sure that solving for the properties of quantum-chromodynamics (QCD) using lattice methods we will capture all properties of the vacuum state? After all, it is hard not to concede that the SM with its 20+ parameters is an effective model. Somewhere in the wealth of information there could, indeed there should, be knowledge hiding that takes us beyond the present day paradigm. 

In my personal interpretation the above words define Andrzej as a physicist. He has devoted a vast majority of his research effort to characterizing subtle correlation and fluctuation features of particle production. This rich physics context will be covered in detail by other contributors. My task is to consider the strong interaction heavy ion \lq knowledge frontiers\rq\ of the present day, and to extrapolate from the present position to the future.

In section \ref{Sec2} I look at the ideas and realities about the formation of quark-gluon plasma in heavy ion collisions, addressing in a personal perspective two specific aspects, in subsection \ref{Sec21} the onset of QGP formation and in subsection \ref{Sec22} the stopping of energy that exceeds expectations. In section \ref{Sec3} I connect the laboratory study of QGP to the physics of the early Universe and address matter-antimatter asymmetry. This is followed in section \ref{Sec4} by a survey of other unsolved fundamental problems that QGP experiments are tangent to. The closing section \ref{Sec5} restates the key frontier questions of this article. 

%%%%%%%%%%%%%%%%%%%%%%%%%%%%%
\begin{figure}[htb]
\centerline{%width=12.5cm
\includegraphics[width=\linewidth]{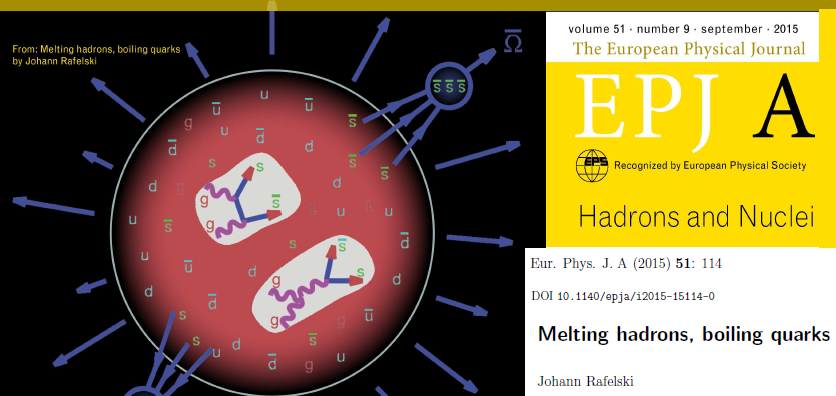}
}
\caption{Illustration of hadronization formation of $\overline{\Omega}(\bar s\bar s\bar s)$ following on formation of strangeness within QGP blob, from the cover of Ref.\,\cite{Rafelski:2015cxa}.}
\label{EPJAcover}
\end{figure}
%%%%%%%%%%%%%%%%%%%%%%%%%%%%%

\section{Heavy-ions and the formation of quark-gluon plasma}\label{Sec2}
%%%%%%%%%%%%%%%%%%%%%%%%%%%%%%%%%%%%%%
\subsection{Deconfinement and strangeness}\label{Sec21}
At two press conferences by CERN and BNL experimental groups the discovery of a new state of matter, the quark-gluon plasma (QGP), was announced more than a decade ago. I have recently chronicled these events~\cite{Rafelski:2015cxa}, see \rf{EPJAcover}. Today at the large-hadron collider (LHC), the properties of QGP are measured against these initial benchmark results. Unlike the finding of a new particle, QGP discovery was a paradigm shift arising when considering a combination of theoretical model studies with numerous relativistic heavy ion (RHI) collisions experimental results obtained at different accelerators and collision energies. 

The identification of the new state of matter relies on particle production patterns. My interest has centered on the understanding the imprint of properties of QGP on the final particle state. Of greatest interest in this context are particles comprising heavier flavor not present in colliding matter, and especially the antimatter version. Since the production of partons that form the final hadron predates the hadron formation process, high abundance of particles can be produced as is illustrated in \rf{EPJAcover}. There is a large abundance of strange antibaryons reported in all heavy-ion experiments, which agrees with the theoretical expectations. 

The more complete story about strange antibaryons is told in Ref.\,\cite{Rafelski:2015cxa}. However, it is important to remember here that since 1980 I emphasize as an interesting and characteristic signal of QGP the ratio $\overline\Lambda/\bar p\approx 1$ (here $\overline\Lambda$ includes feed from $\overline\Sigma^0\to \overline\Lambda+\gamma$). The anomalous ratio increases as energy decreases, naturally only if QGP is formed. Both the theory and the experiment were reviewed in my Lecture Notes for the Krak\'ow School of Theoretical Physics, XLVIe Course, 2006, Section 2~\cite{Rafelski:2006ny}.

This strange antibaryon observable is also presented as the ratio of the yield in A+A collisions compared to $p$+A or/and $p+p$. As energy is reduced the difficulty of producing any background at all increases and thus such a ratio peaks at the lowest energy at which the collective mechanism of strange antibaryon production shown in \rf{EPJAcover} is operational. 

However, a yet simpler observable is also available. Emanuele Quercigh pointed out to me that he could measure changes in K/$\pi$ ratio at 10\% level, so any effect related to QGP must be larger. I answered he should look at multistrange antibaryons, which he did. The outcome was the discovery of strange antibaryon enhancement that remains to date the largest medium effect in heavy ion collisions, as large as a factor 20. The summary of the current results is shown in \rf{AntiB}. 

%%%%%%%%%%%%%%%%%%%%%%%%%%%%%
\begin{figure}[htb]
\centerline{%width=12.5cm
\includegraphics[width=0.95\linewidth]{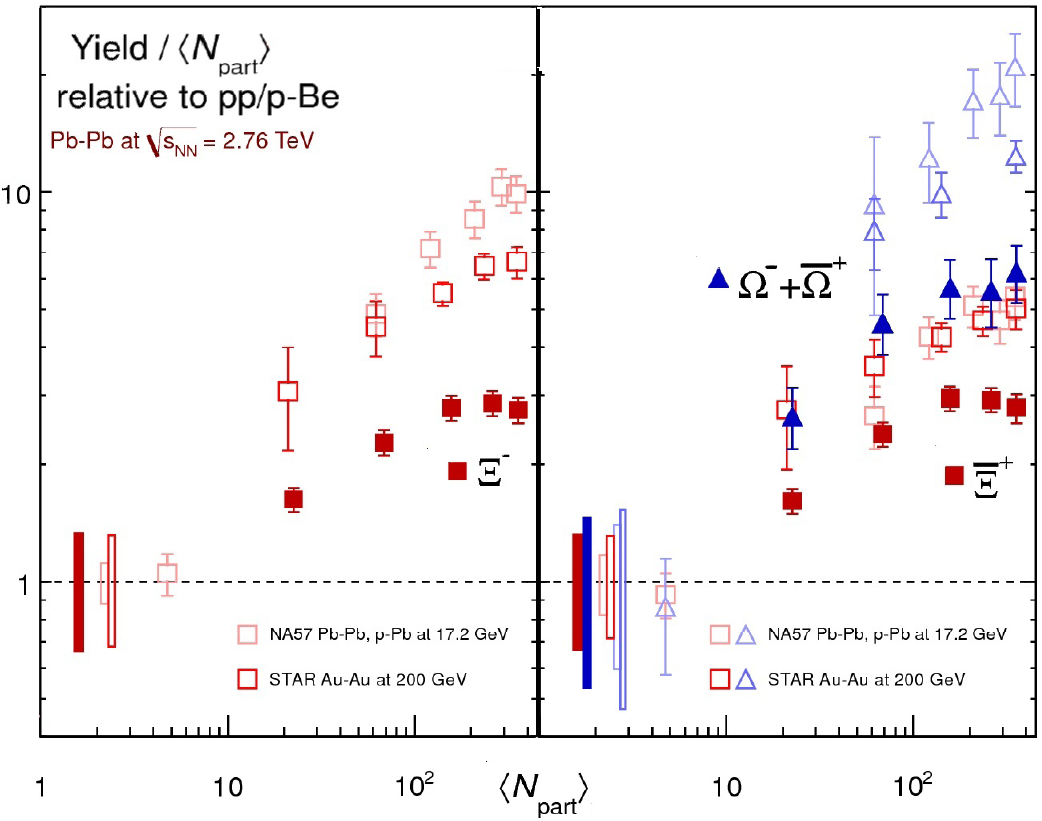}
}
\caption{Strange (anti)baryon enhancement with reference to $pp$ and $pBe$ measured at LHC-ALICE compared to SPS and STAR results.} 
\label{AntiB} 
\end{figure}
%%%%%%%%%%%%%%%%%%%%%%%%%%%%%

%%%%%%%%%%%%%%%%%%%%%%%%%%%%%
\begin{figure}[htb]
\centerline{%width=12.5cm
\includegraphics[width=0.95\linewidth]{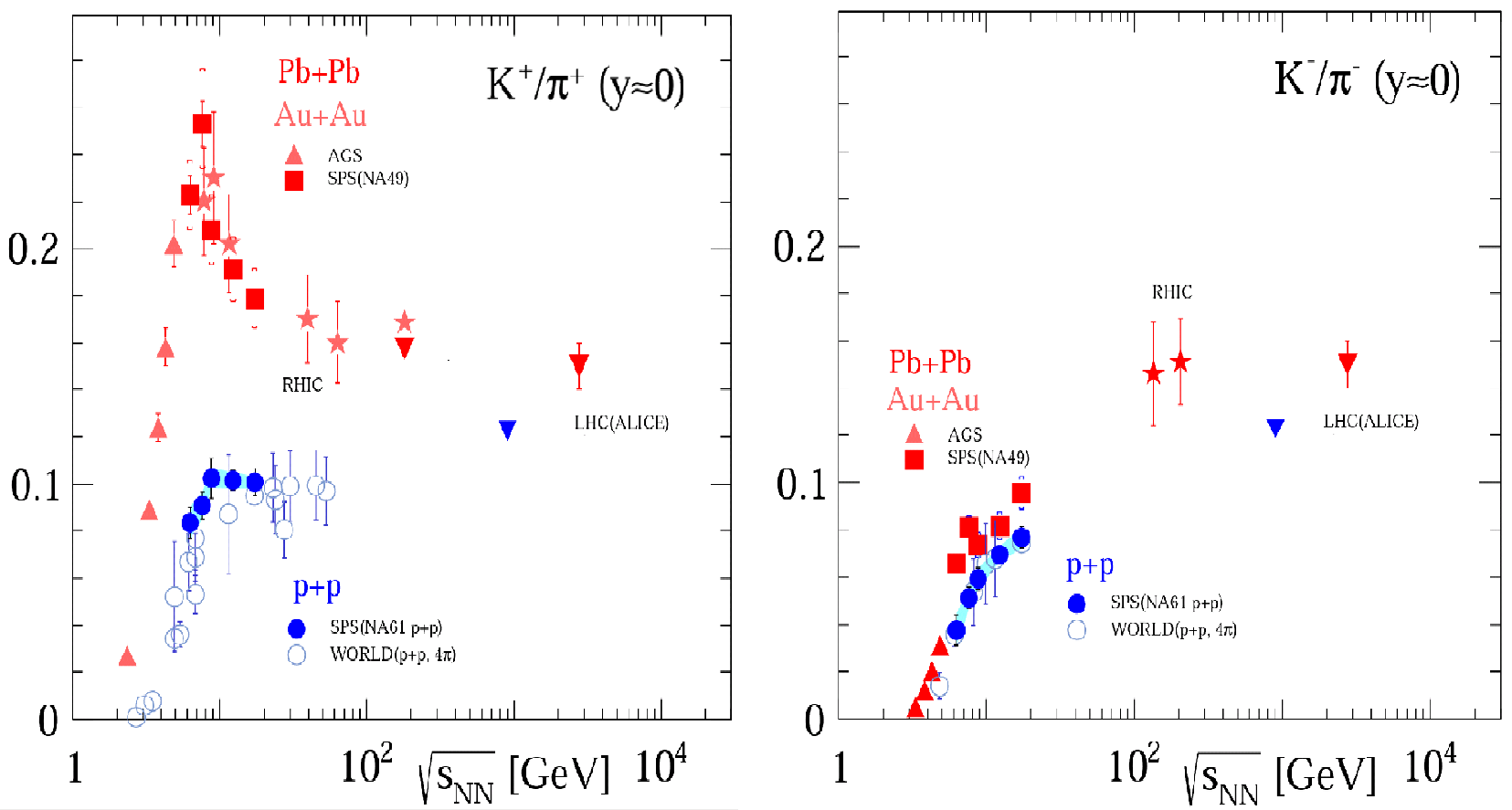}
}
\caption{Update of results presented in Ref.\,\cite{Gazdzicki:2010iv} (M. Ga\'zdzicki, private communication).}
\label{MarekNewHorn} 
\end{figure}
%%%%%%%%%%%%%%%%%%%%%%%%%%%%%
However, a few months after this discussion which took place in Winter 1984/85 I decided to take a second look at Emanuele\rq s question obout K/$\pi$ and I presented K$^+/\pi^+$ as a measure of the ratio of strangeness production to entropy in QGP~\cite{Glendenning:1984ta} (in this early paper the K$^+/\pi^+$ includes only directly produced particles, pions from resonance decay dilute the presented ratio to present day experimental values). In order to discriminate between the confined and deconfined phases of matter the benchmark could be $d$+A which has isospin symmetry similar to A+A.

The discovery~\cite{Gazdzicki:2010iv} of the \lq K$^+/\pi^+$ horn\rq\ shown in \rf{MarekNewHorn} was a watershed event that galvanized interest in the characterization of the onset of deconfinement. Today there is a BES (beam energy scan) program at RHIC, and at CERN the NA61 experiment is taking data in a wide range of energies for many collision systems. All of this is aimed at a) finding the critical point where at a finite value of baryo-chemical potential $\mu_\mathrm{B}$ a true phase transition sets in, and b) identifying the onset of deconfinement. 

The question is open if K$^+/\pi^+$ horn signals a) onset of deconfinement, b) the critical point, or c) is signaling a rapid change in the properties of compressed and excited nuclear matter. This is so since an anomaly in  K$^+/\pi^+$ ratio related to properties of a phase of matter can be confounded with an anomaly related to collision dynamics. I have little doubt that in the coming decade an answer about the properties of QGP phase boundary will be sought and obtained. One can say that we stand today at this frontier, waiting to cross.

\subsection{The McLerran-Bjorken transparency}\label{Sec22}
%%%%%%%%%%%%%%%%%%%%%%%%%%%%%
In 1980 Larry McLerran and collaborators proposed, based on a fragmentation view of $pp$ reactions, a two-fireball model~\cite{Anishetty:1980zp} of nuclear collisions. Citing from abstract: \lq\lq We discuss central collisions between heavy nuclei of equal baryon number at extremely high energies\ldots fragmentation-region fragments thermalize, \ldots discuss the possible formation of hot, dense quark plasmas in the fireballs.\rq\rq\ In this approach there are few if any particles in the central rapidity region in contrast to models developed in that time period by Hagedorn and myself~\cite{Hagedorn:1980kb,Rafelski:2016hnq}. 

%%%%%%%%%%%%%%%%%%%%%%%%%%%%%
\begin{figure}[htb]
\centerline{%width=12.5cm
\includegraphics[angle=-90,width=0.47\linewidth]{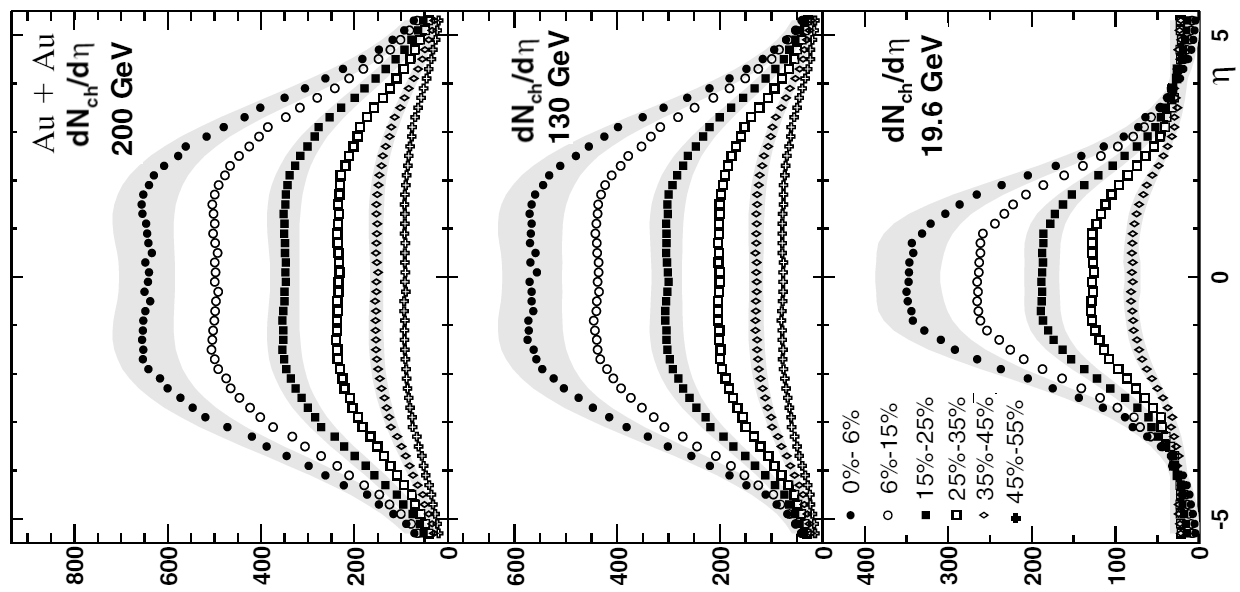}
\includegraphics[angle=-90,width=0.49\linewidth]{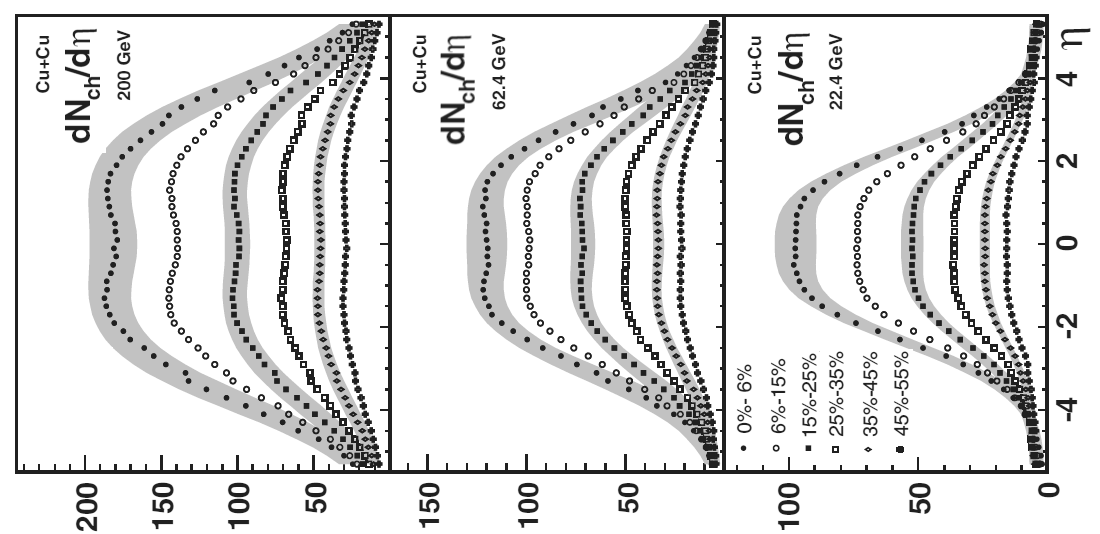}
}
\caption{RHIC-PHOBOS~\cite{Back:2002wb,Alver:2007aa} rapidity distributions of charged particles as a function of psudorapidity $\eta$ for three collision energies and a range of centralities as indicated, on left for Au+Au~\cite{Back:2002wb} and on right Cu+Cu~\cite{Alver:2007aa}.}
\label{PhobosFig}
\end{figure}
%%%%%%%%%%%%%%%%%%%%%%%%%%%%%

There is no sign of two projectile-target fragmentation fireballs in the RHIC-PHOBOS multiplicity results shown in \rf{PhobosFig} either in Au+Au~\cite{Back:2002wb}, or the lighter Cu+Cu~\cite{Alver:2007aa} collisions systems in the wide range of energies displayed. In fact Larry\rq s model was very soon complemented by work which addressed QGP formation in the central rapidity region~\cite{Bjorken:1982qr}. Bjorken proposed rapidity scaling: when the energy of colliding nuclei is high enough, there would need to be a flat \lq central\rq\ rapidity region as it is impossible to distinguish one rapidity value from another, the nuclear fragments having left, with energy trailing their departure. To understand this physics idea, think of an airplane track in the sky (or a charged particle track in a bubble chamber); what we see is energy deposited per unit of distance which does not depend that much on where the airplane is when we watch the sky. 

Numerical work~\cite{Baym:1984sr} in 1+3 dimensions including transverse expansion followed and continues to this day, but I do not believe that anyone has come close to reproducing the charged particle multiplicity results seen in \rf{PhobosFig}. In the formal QGP discovery PHOBOS report~\cite{Back:2004je}, in the legend to Figure 23 showing true pion rapidity distributions drawn from experiments E895 at AGS, NA49 at SPS and BRAHMS at RHIC, I read \lq\lq ($\pi^+$ rapidity) yields in rapidity space are well represented by Gaussians with no evidence for a broad midrapidity plateau\rq\rq. Along with the PHOBOS charged particle multiplicity data this result invalidates the parton transparency ideas of McLerran and Bjorken up and including the top RHIC energy, for which they were developed.

What high energy means, and when as a function of energy onset of rapidity scaling begins, can be subject to debate. Turning now to LHC where today experiments operate at $\sqrt{s_\mathrm{NN}}=5.02$\,TeV -- that is 25 times the RHIC energy maximum, I note that we do not have any data on charged particle multiplicity with full rapidity coverage. In the rapidity coverage available based on transverse energy distribution there is an inkling of a peaked distribution for $y\to 0$ but as it would be not signed by the CMS experimental group, I cannot present a citation. 

However, we now have a remarkable result showing the charged particle multiplicity at central rapidity~\cite{Adam:2015ptt} in A-A collisions up to $\sqrt{s_\mathrm{NN}}=5.02$\,TeV. The yield rises as $(\sqrt{s_\mathrm{NN}})^{0.310(8)}$, while in $pp$, $p$A and $d$ the rise is $(\sqrt{s_\mathrm{NN}})^{0.206(4)}$. ALICE collaboration amplifies in text: \lq\lq \ldots (ALICE) $p$-Pb and PHOBOS for $d$-Au collisions fall on the curve for proton-proton collisions, indicating that the strong rise in AA (multiplicity at central rapidity) is not solely related to the multiple collisions undergone by the participants, since the proton in $p$-A collisions also encounters multiple nucleons.\rq\rq\ 

The rapid growth of central rapidity charged particle multiplicity with energy in A-A collisions is in my opinion adding further evidence that at LHC like at RHIC there is a peak in multiplicity distribution as a function of rapidity and thus we observe concentration of particles when taking data at $\eta=0$. I conclude: after 25 years of experiments at ever higher energy reaching now to collision of partons at TeV energy scale, the rapidity scaling, a trademark picture of partons passing each other leaving cooked nuclei behind and pulling a long trail of energy~\cite{Anishetty:1980zp,Bjorken:1982qr} has {\bf not} been found. Many are working, seeking an answer to this big riddle, which we will need to understand in the coming decade.

Our ignorance about reaction mechanisms that cause the formation of high entropy=particle multiplicity density in the central rapidity region does not affect the physics of quark-gluon plasma that we derive from final state hadron abundances. This is so since the reaction mechanisms at LHC occur at an energy scale of 1000 GeV per parton, while the soft QGP physics occurs at parton energy of $3T=0.5$ GeV per parton, that is after initial partons have softened in energy by a factor 1000. 
%%%%%%%%%%%%%%%%

%%%%%%%%%%%%%%%%%%
\section{The primordial quark universe in laboratory}\label{Sec3}
%%%%%%%%%%%%%%%%%%%%%%%%%%%%%%%%
\subsection{Big-bang and micro-bang time scales}
One of the original motivations for the development of the relativistic heavy ion collision program is the recreation and study of the extreme temperature conditions prevailing in the early Universe in presence of the quark-gluon phase of matter. This era of evolution began at about 10\,ns after the big bang, and lasted through the time when QGP froze into individual hadrons at about 18\,$\mu$s. The 10\,ns is a rough estimate of when the electro-weak symmetry breaking occurred creating the format of (effective) elementary interactions within the SM we are familiar with today.

The above two time scales of Universe evolution, {\it i.e.\/} ns and $\mu$s, are greatly larger as compared to the natural time scale that governs the laboratory collision processes. Taking $R=10$\,fm to be the natural distance scale of processes governing QGP and considering relativistic particle speed, the lifespan of QGP made in the laboratory is $\tau=3\times 10^{-23}$\,s. This implies that the laboratory study of the early Universe must be supplemented with theoretical modeling to accommodate the connection to laboratory experiments.

%%%%%%%%%%%%%%%%%%%%%%%%%%%%%
\subsection{Matter-antimatter symmetry}
Although around us today the world is made of baryons, in the QGP phase the Universe is nearly matter-antimatter symmetric with light $u,d$, strange and some heavier quarks and antiquarks present in practically equal and very large abundance. As the expansion cooling of the Universe takes the matter below deconfinement conditions, the early Universe evolved across the matter-antimatter annihilation era, leaving behind a tiny $10^{−9}$ residual matter asymmetry fraction. 

The situation is very different in the laboratory big-bang recreation experiment since the freshly created QGP has lots of empty space to explosively expand into, and thus matter and antimatter begin to free-stream without any significant depletion of their abundance, laboratory generated QGP is a source of antimatter. The antimatter formation by way of QGP formation and the potential for its harvesting in RHI collisions is a possible future application of the present day foundational RHI physics.

Because we collide baryonic matter, we expect that the QGP we form is weighted in its content towards baryons, leading in LHC experiments to a stronger asymmetry than governed the early Universe. Said differently, even if the colliding quarks in nuclei fly out from the fireball maintaining nearly their original rapidity, some net baryon density could remain in the central fireball. This is the second difference we note between QGP that filled the Universe and the laboratory effort: in the early Universe the baryo-chemical potential $\mu_\mathrm{B}$ is measured in terms of a fraction of eV~\cite{Fromerth:2002wb}, but current thinking says that $\mu_\mathrm{B}$ that remains in central rapidity in LHC experiments is at a fraction of MeV or larger. However, the determination of the value of $\mu_\mathrm{B}$ at LHC requires a dedicated effort and is pending.

I think it is not at all guaranteed that at the large hadron collider (LHC) reaction energy any baryon excess in central fireball is created by matter \lq stopping\rq. One must ask, is this baryon density due to the matter we bring into collision or due to some new process that in the end produces an excess of baryon number, be it in laboratory, be it in the early Universe? 
 
\subsection{Searching the origin of baryon number}
%%%%%%%%%%%%%%%%%%%%%%%%%%%%%%%%%%%%%
The question about the origin of baryon asymmetry in LHC energy scale heavy ion collisions addresses the unresolved riddles about the origin of baryon asymmetry in the Universe: 
1) What is the mechanism and interaction that creates the baryon asymmetry; 
2) In which time era of the Universe did the present day net baryon number surrounding us originate? 
The reason that these questions fascinate is that they directly and experimentally confront the questions about stability of matter and what is mass of matter. 

Most believe that the net baryon asymmetry that we see in the Universe is not due to an initial condition pre-established at the onset of the big bang. The big question is, if in the environment of QGP forming RHI collisions, we should invest effort into looking for the development of baryon asymmetry. This may be a long shot since the laboratory collision time scale is very short compared to the Universe. On the other hand, the conditions are sufficiently different and experimental tools allowing the measurement of such an asymmetry are relatively easily accessible for considering this question seriously.
 
For baryon number to appear in our space-time domain of the Universe, the three Sakharov conditions have to be fulfilled. I compare now their contents with what RHI collisions can provide:\\

{\bf 1)} There must be non-equilibrium: the Universe cannot evolve completely in equilibrium, or else whatever creates the baryon asymmetry will also undo it. This requirement is generally understood to mean that the asymmetry has to originate in the period of a phase transformation, and the focus of attention has been on electro-weak symmetry restoring condition at a temperature scale $1000\times \Th$ above the hadronization of QGP. 

The time available for the asymmetry to arise is in this condition on the scale of $10^{-8}$\,s. One must note that in laboratory experiments involving QGP phase there is much more nonequilibrium compared to the Universe evolution which, given the long time constant, should be very smooth across the phase transformation from quarks to hadrons. Therefore it is possible that in laboratory experiments QGP hadronization that is far from equilibrium is a suitable environment allowing mechanisms of baryon asymmetry formation.\\

{\bf 2)} During the nonequilibrium time period the interactions must be able to differentiate between matter and antimatter, or else how could the residual asymmetry be matter dominated? This requires at least CP symmetry breaking -- CPT breaking, e.g. difference of mass between particles and antiparticles will also do. In the standard model of particle physics the complex phase of the Kobayashi-Maskawa flavor mixing breaks CP -- the breaking is said to be too weak to produce the asymmetry we see. 

However, this discussion pinpoints flavor mixing as the origin of CP non-conservation. Quark-gluon plasma created in laboratory is full of strangeness, quark flavor of second family. Current experiments are exploring charm content, completing the understanding of the second flavor family in QGP. Future efforts should also allow access to bottom quarks from the third particle family. Unlike in the adiabatically evolving quark Universe, in the laboratory we have all these three quark flavor families present in chemical nonequilibrium.\\

{\bf 3)} If and when in some space-time domain of the Universe excess of baryons over antibaryons is to arise there must be a baryon number conservation violating process that produces the baryon asymmetry needed. There are several ways to achieve this: i) an outright destruction of baryon number preserving color charge, electrical charge in a $B-L$ (baryon minus lepton number) preserving process of the grand unified theory (GUT) type, where the color $\bar 3$ charge of diquark is indicated below
\begin{equation}
(uu)_{\bar 3}\to \bar d + l^+\ \Leftrightarrow\ p\to \pi^0+ e^+\;;
\end{equation}
ii) a force that has baryon number as charge, and thus can separate baryons from antibaryons, creating local baryon asymmetry without altering overall baryon content -- such a force would need to be short ranged, and/or weak, or else it would have been discovered already; iii) a space-time anomaly where quarks hide their color charge turning in the process into e.g. leptons. i) has not been observed in experimental searches for proton decay at the level that would be required and consistent with the standard model. ii) A baryon number based force could be more effective in creating in heavy ion collisions baryon asymmetry acting at a lower temperature in laboratory environment when compared to electro-weak transition era in the early Universe. iii) requires activation energy and may not be accessible in RHI collisions. While only ii) may be accessible in RHI experiments it is certainly possible that a mix of all three mechanisms is present in the Universe, and there could be another mechanism that has not been imagined above which is RHI accessible -- such that an effect can be observed. 

An example of experimental signature of baryon asymmetry that one wants to look for in the central rapidity region is an excess abundance of exotic antibaryons comprising quarks of second and third families only, beginning with the search for asymmetry in the $\overline{\Omega}(\bar s\bar s\bar s)$ excess over ${\Omega}( s s s)$ in central rapidity region at LHC and extending this to $\overline{\Omega}_c(\bar c \bar s\bar s)/{\Omega}_c( c s s)$ and carrying on to every exotic state involving the third family such $\overline{\Omega}_{bc}(\bar b \bar c\bar s)/{\Omega_{bc}}( b c s)$. The reason that we do not want to look at first family $u,d$ baryons (protons) is that many are brought into collision and thus there is a natural bias to observe more baryons than antibaryons even at the most central rapidity region $y=0$. The reason that antibaryon asymmetry could favor anti-exotic baryons in central rapidity domain is that the baryon number shooting through into projectile and target rapidity region should act as an attractor also for baryon number locked in exotic flavor baryons.
%%%%%%%%%%%%%%%%%%

\section{Other foundational frontiers}\label{Sec4}
\subsection{Seeking the origin of entropy}
%%%%%%%%%%%%%%%%%%%%%%%%%%%%%%%%%%%%%
A long-standing question, formulated before first RHI collision results came to be, is: How does it happen that in collisions of relatively small in size nuclei the enormous amount of inelasticity and entropy can be created in the time available for the formation of a thermal quark-gluon plasma state? We can easily formulate a computational answer based on initial state chaoticity, scattering, and parton splitting reactions. However, this conventional approach imposes classical processes on a system that cannot be classical in this initial stage. The question thus is how to use quantum theory to describe the formation of a thermal QGP. An answer would advance not only the understanding of the first primordial time instant in RHI collision, but in my opinion, it would add to our comprehension of quantum physics.

\subsection{What is mass?}
%%%%%%%%%%%%%%%%%%%%%%%%%%%%%%%%%%%%%
The question: What is mass and how does it originate is studied in RHI collisions by melting hadrons into quarks. Mass of matter originates in quarks, which are confined and are not freely roaming. The color charge of quarks is at the heart of the phenomenon. This charge needs to find a counter charge. The ability of three red-green-blue quarks to turn into a colorless baryon or antibaryon is providing us with stable baryonic matter. The question that begs attention is how when massive particles emerge from the quark-gluon soup such that on \lq balance\rq\ baryon number and other discrete quantum numbers (charge, angular momentum, strangeness and other flavors) balance such that particles emerge with correct characteristics.

The QGP hadronization process contains this key information about how energy turns into mass of well balanced particles produced. Something keeps track of all degrees of freedom assuring validity of mass based models of particle production according to statistical physics principles. I am pretty sure that beyond the equations we use, deeper insights lurk, which will surface in due time as we keep up the process of inquiry.

\subsection{What is flavor?}
%%%%%%%%%%%%%%%%%%%%%%%%%%%%%%%%%%%%%
What is flavor? In elementary particle collisions, we deal with a few, and in most cases only one, pair of newly created second, or third flavor family of particles at a time. A new situation arises in the QGP formed in relativistic heavy ion collisions. QGP includes a large number of particles from the second family: the strange quarks and also, the yet heavier charmed quarks; and from the third family at the LHC we expect an appreciable abundance of bottom quarks. The novel ability to study a large number of these second and third generation particles offers a new opportunity to approach the riddle of flavor in an experiment.

\subsection{Matter stability}
%%%%%%%%%%%%%%%%%%%%%%%%%%%%%%%%%%%%%
In relativistic heavy ion collisions the kinetic energy of ions feeds the growth of quark population. These quarks ultimately turn into final state material particles. This means that we study experimentally the mechanisms leading to the conversion of the colliding ion kinetic energy into, flavor dependent, mass of matter. Does the study of how kinetic energy turns into mass teach anything about how it is possible to convert matter into energy in the laboratory?
 
%%%%%%%%%%%%%%%%%%
\section{All questions}\label{Sec5}
I compile here a short but wide-ranging list of frontier questions that have been introduced in the text:
\begin{enumerate}
\item 
The SM with its 20+ parameters is an effective model and that includes QCD: Are we sure that solving for the properties of quantum-chromodynamics (QCD) using lattice methods we will capture all properties of the vacuum state?
\item Does K$^+/\pi^+$ horn signal a) onset of deconfinement, b) an accidental stumble onto the critical point, or c) a rapid change in the properties of compressed and excited nuclear matter?
\item Why is there non-transparency in A+A collisions at highest energies? 
\item At the large hadron collider (LHC) energy scale, is the baryon excess in central fireball 
created or scattered from matter we bring into collision? 
\item Is there a chance that we can discover a baryon number violating process in LHC energy scale heavy ion collisions?
\item How does the initial state decohere allowing the enormous amount of entropy to be created in the time available for the formation of a thermal quark-gluon plasma state?
\item How does baryon number and other discrete quantum numbers balance and all particles emerge with correct characteristics in freezing of QGP into hadrons?
\item What is flavor? QGP is the only experimental environment that includes a large number of particles from the second flavor family and some from the third, offering a new opportunity to approach the riddle of flavor in an experiment.
\item One can wonder aloud if the study of how kinetic energy turns into mass teaches how it is possible to convert matter into energy in the laboratory?
\end{enumerate}

%%%%%%%%%%%%%%%%%%%%%%%%%%%%%%%%%%%%%%%%%%%%%%%%%


\begin{thebibliography}{99}
%\cite{Bialas:1988wc}
\bibitem{Bialas:1988wc}
A.~ Bia\l{}as and R.~B.~Peschanski,
\lq\lq Intermitency in Multiparticle Production at High-Energy,\rq\rq\
Nucl.\ Phys.\ B {\bf 308}, 857 (1988),
\url{doi:10.1016/0550-3213(88)90131-9}.
 

%\cite{Rafelski:1996hf}
\bibitem{Rafelski:1996hf} 
 J.~Rafelski, J.~Letessier and A.~Tounsi,
\lq\lq Strange particles from dense hadronic matter,\rq\rq\
Acta Phys.\ Polon.\ B {\bf 27}, 1037 (1996),
\url{http://www.actaphys.uj.edu.pl/fulltext?series=Reg&vol=27&page=1037}.
%[nucl-th/0209080]. 

%\cite{Bialas:1998ea}
\bibitem{Bialas:1998ea} 
 A.~ Bia\l{}as,
 \lq\lq Quark model and strange baryon production in heavy ion collisions,\rq\rq
 Phys.\ Lett.\ B {\bf 442}, 449 (1998),
 \url{doi:10.1016/S0370-2693(98)01250-7}.
% [hep-ph/9808434].

%\cite{Torrieri:2004zz}
\bibitem{Torrieri:2004zz} 
 G.~Torrieri, S.~Steinke, W.~Broniowski, W.~Florkowski, J.~Letessier and J.~Rafelski,
 \lq\lq SHARE: Statistical hadronization with resonances,\rq\rq\
 Comput.\ Phys.\ Commun.\ {\bf 167}, 229 (2005),
 \url{doi:10.1016/j.cpc.2005.01.004};\\
% [nucl-th/0404083].
%\cite{Torrieri:2006xi}
%\bibitem{Torrieri:2006xi} 
 G.~Torrieri, S.~Jeon, J.~Letessier and J.~Rafelski,
 \lq\lq SHAREv2: Fluctuations and a comprehensive treatment of decay feed-down,\rq\rq\
 Comput.\ Phys.\ Commun.\ {\bf 175}, 635 (2006),
 \url{doi:10.1016/j.cpc.2006.07.010};\\
% [nucl-th/0603026].
 %%CITATION = doi:10.1016/j.cpc.2006.07.010;%%
 %59 citations counted in INSPIRE as of 13 Jun 2016
%\cite{Petran:2013dva}
%\bibitem{Petran:2013dva} 
 M.~Petran, J.~Letessier, J.~Rafelski and G.~Torrieri,
 \lq\lq SHARE with CHARM,\rq\rq\
 Comput.\ Phys.\ Commun.\ {\bf 185}, 2056 (2014),
 \url{doi:10.1016/j.cpc.2014.02.026}.
% [arXiv:1310.5108 [hep-ph]].
 

%\cite{Bialas:2005ij}
\bibitem{Bialas:2005ij} 
 A.~ Bia\l{}as and J.~Rafelski,
 \lq\lq Balance of baryon number in the quark coalescence model,\rq\rq
 Phys.\ Lett.\ B {\bf 633}, 488 (2006),
 \url{doi:10.1016/j.physletb.2005.11.084}.
% [hep-ph/0508084].
 
%\cite{Rafelski:2015cxa}
\bibitem{Rafelski:2015cxa} 
 J.~Rafelski,
 \lq\lq Melting Hadrons, Boiling Quarks,\rq\rq\
 Eur.\ Phys.\ J.\ A {\bf 51}, 114 (2015)
 \url{doi:10.1140/epja/i2015-15114-0}
% [arXiv:1508.03260 [nucl-th]].

%\cite{Rafelski:2006ny}
\bibitem{Rafelski:2006ny} 
 J.~Rafelski and J.~Letessier,
 \lq\lq Status of Strangeness-Flavor Signature of QGP,\rq\rq\
 Acta Phys.\ Polon.\ B {\bf 37}, 3315 (2006),
 \url{http://www.actaphys.uj.edu.pl/fulltext?series=Reg&vol=37&page=3315}.
% [hep-ph/0610106].

%\cite{Gazdzicki:2010iv}
\bibitem{Gazdzicki:2010iv} 
 M.~Gazdzicki, M.~Gorenstein and P.~Seyboth,
 \lq\lq Onset of deconfinement in nucleus-nucleus collisions: Review for pedestrians and experts,\rq\rq\
 Acta Phys.\ Polon.\ B {\bf 42}, 307 (2011),
 \url{doi:10.5506/APhysPolB.42.307}.
% [arXiv:1006.1765 [hep-ph]].
 
%\cite{Glendenning:1984ta}
\bibitem{Glendenning:1984ta} 
 N.~K.~Glendenning and J.~Rafelski,
 \lq\lq Kaons and Quark Gluon Plasma,\rq\rq
 Phys.\ Rev.\ C {\bf 31}, 823 (1985),
 \url{doi:10.1103/PhysRevC.31.823}.

%\cite{Anishetty:1980zp}
\bibitem{Anishetty:1980zp} 
 R.~Anishetty, P.~Koehler and L.~D.~McLerran,
 \lq\lq Central collisions between heavy nuclei at extremely high-energies: The fragmentation region,\rq\rq\
 Phys.\ Rev.\ D {\bf 22}, 2793 (1980);
\url{doi:10.1103/PhysRevD.22.2793}; 
and: L. McLerran: a) in Proceedings of the \lq\lq 5th High Energy Heavy Ion Study,\rq\rq\ Berkeley, California, 1981; b) invited talk presented at Physics in Collision: High Energy ee/ep/pp Interactions, Blacksburg, Virginia, 1981.
 
%\cite{Hagedorn:1980kb}
\bibitem{Hagedorn:1980kb} 
 R.~Hagedorn and J.~Rafelski,
 \lq\lq Hot Hadronic Matter and Nuclear Collisions,\rq\rq\
 Phys.\ Lett.\ B {\bf 97}, 136 (1980),
 \url{doi:10.1016/0370-2693(80)90566-3}.
 
%\cite{Rafelski:2016hnq}
\bibitem{Rafelski:2016hnq} 
 J.~Rafelski, Editor,
 \lq\lq Melting Hadrons, Boiling Quarks - From Hagedorn Temperature to Ultra-Relativistic Heavy-Ion Collisions at CERN : With a Tribute to Rolf Hagedorn,\rq\rq\
 \url{doi:10.1007/978-3-319-17545-4} (Springer Open 2016).
 %%CITATION = doi:10.1007/978-3-319-17545-4;%%

%\cite{Back:2002wb}{Alver:2007aa}
\bibitem{Back:2002wb} 
 B.~B.~Back {\it et al.}, [PHOBOS-Collaboration],
 \lq\lq The Significance of the fragmentation region in ultrarelativistic heavy ion collisions,\rq\rq\
 Phys.\ Rev.\ Lett.\ {\bf 91}, 052303 (2003)
\url{doi:10.1103/PhysRevLett.91.052303}.
% [nucl-ex/0210015].

%\cite{Alver:2007aa}
\bibitem{Alver:2007aa} 
 B.~Alver {\it et al.}, [PHOBOS-Collaboration],
 \lq\lq System size, energy and centrality dependence of pseudorapidity distributions of charged particles in relativistic heavy ion collisions,\rq\rq\
 Phys.\ Rev.\ Lett.\ {\bf 102}, 142301 (2009),
\url{doi:10.1103/PhysRevLett.102.142301}.
% [arXiv:0709.4008 [nucl-ex]].
 
%\cite{Bjorken:1982qr}
\bibitem{Bjorken:1982qr} 
 J.~D.~Bjorken,
 \lq\lq Highly relativistic nucleus-nucleus collisions: The central rapidity region,\rq\rq\
 Phys.\ Rev.\ D {\bf 27}, 140 (1983),
 \url{doi:10.1103/PhysRevD.27.140}.
 
%\cite{Baym:1984sr}
\bibitem{Baym:1984sr} 
 G.~Baym, B.~L.~Friman, J.~P.~Blaizot, M.~Soyeur and W.~Czyz,
 \lq\lq Hydrodynamics of ultrarelativistic heavy ion collisions,\rq\rq\
 Nucl.\ Phys.\ A {\bf 407}, 541 (1983),
 \url{doi:10.1016/0375-9474(83)90666-8}.
 
%\cite{Back:2004je}
\bibitem{Back:2004je} 
 B.~B.~Back {\it et al.}, [PHOBOS-Collaboration],
 \lq\lq The PHOBOS perspective on discoveries at RHIC,\rq\rq\
 Nucl.\ Phys.\ A {\bf 757}, 28 (2005),
 \url{doi:10.1016/j.nuclphysa.2005.03.084}.
% [nucl-ex/0410022].
 
%\cite{Adam:2015ptt}
\bibitem{Adam:2015ptt} 
 J.~Adam {\it et al.} [ALICE Collaboration],
 \lq\lq Centrality dependence of the charged-particle multiplicity density at midrapidity in Pb-Pb collisions at $\sqrt{s_{\rm NN}}$ = 5.02 TeV,\rq\rq\
 Phys.\ Rev.\ Lett.\ {\bf 116}, 222302 (2016),
 \url{doi:10.1103/PhysRevLett.116.222302}.
% [arXiv:1512.06104 [nucl-ex]].

%\cite{Fromerth:2002wb}
\bibitem{Fromerth:2002wb} 
 M.~J.~Fromerth and J.~Rafelski,
 \lq\lq Hadronization of the quark Universe,\rq\rq\ 
\url{http://arxiv.org/abs/astro-ph/0211346};
 astro-ph/0211346. These original results can be found in 
 %\cite{Fromerth:2012fe}
%\bibitem{Fromerth:2012fe} 
 M.~J.~Fromerth, I.~Kuznetsova, L.~Labun, J.~Letessier and J.~Rafelski,
 \lq\lq From Quark-Gluon Universe to Neutrino Decoupling,\rq\rq\
 Acta Phys.\ Polon.\ B {\bf 43}, 2261 (2012),
 \url{doi:10.5506/APhysPolB.43.2261}.
% [arXiv:1211.4297 [nucl-th]].

\end{thebibliography}
\end{document}